\documentclass[useAMS,usenatbib]{mn2e}
\usepackage{graphicx}

\title[The Scale of the Problem]{The Scale of the Problem : Recovering Images of Reionization with GMCA}
\author[E. Chapman et al.]
{Emma Chapman,$^1$\thanks{eow@star.ucl.ac.uk}
 Filipe B. Abdalla,$^1$ J. Bobin,$^2$ J.-L. Starck, $^2$ Geraint Harker,$^{3,4}$ 
\newauthor
Vibor Jeli\'{c},$^{5}$ Panagiotis Labropoulos,$^{5,6}$ Saleem Zaroubi,$^6$ Michiel A. Brentjens, $^5$ 
\newauthor 
A.G. de Bruyn, $^{5,6}$ L.V.E. Koopmans$^6$  \\ 
$^1$Department of Physics \& Astronomy, University College London, Gower Street, London, WC1E 6BT \\
$^2$ Service d'Astrophysique (DAPNIA/SEDI-SAP), Centre Europeen d'Astronomie/Saclay, F-91191 Gif-sur-Yvette Cedex France\\
$^3$Center for Astrophysics and Space Astronomy, 389 UCB, University of Colorado, Boulder, CO 80309-0389, USA \\
$^4$NASA Lunar Science Institute, NASA Ames Research Center, Moffett Field, CA 94035, USA \\
$^5$ASTRON, PO Box 2, NL-7990AA Dwingeloo, the Netherlands\\
$^6$Kapteyn Astronomical Institute, University of Groningen, PO Box 800, 9700AV Groningen, the Netherlands }

\def\LaTeX{L\kern-.36em\raise.3ex\hbox{a}\kern-.15em
    T\kern-.1667em\lower.7ex\hbox{E}\kern-.125emX}

\begin{document}

\maketitle

\begin{abstract}
The accurate and precise removal of 21-cm foregrounds from Epoch of Reionization redshifted 21-cm emission data is essential if we are to gain insight into an unexplored cosmological era. We apply a non-parametric technique, Generalized Morphological Component Analysis or \textsc{gmca}, to simulated LOFAR-EoR data and show that it has the ability to clean the foregrounds with high accuracy. We recover the 21-cm 1D, 2D and 3D power spectra with high accuracy across an impressive range of frequencies and scales. We show that \textsc{gmca} preserves the 21-cm phase information, especially when the smallest spatial scale data is discarded. While it has been shown that LOFAR-EoR image recovery is theoretically possible using image smoothing, we add that wavelet decomposition is an efficient way of recovering 21-cm signal maps to the same or greater order of accuracy with more flexibility. By comparing the \textsc{gmca} output residual maps (equal to the noise, 21-cm signal and any foreground fitting errors) with the 21-cm maps at one frequency and discarding the smaller wavelet scale information, we find a correlation coefficient of 0.689, compared to 0.588 for the equivalently smoothed image. Considering only the pixels in a central patch covering 50$\%$ of the total map area, these coefficients improve to 0.905 and 0.605 respectively and we conclude that wavelet decomposition is a significantly more powerful method to denoise reconstructed 21-cm maps than smoothing.
\end{abstract}


\section{Introduction}
When the first ionizing sources appeared 400 million years after the Big Bang, the Universe emerged from the `Dark Ages' and began to be reionized. This Epoch of Reionization (EoR) is on the verge of being directly observed for the first time, with a new generation of radio telescopes beginning to see first light (e.g. Low Frequency Array (LOFAR)\footnote{http://www.lofar.org/} (van Haarlem et al., in preparation), Giant Metrewave Radio Telescope (GMRT)\footnote{http://gmrt.ncra.tifr.res.in/}, Murchison Widefield Array (MWA)\footnote{http://www.mwatelescope.org/},Precision Array to Probe the Epoch of Reionization (PAPER)\footnote{http://astro.berkeley.edu/~dbacker/eor/}, 21 Centimeter Array (21CMA)\footnote{http://21cma.bao.ac.cn/}).

The vast majority of EoR observations will take advantage of the 21-cm spectral line - produced by a spin flip in neutral hydrogen \citep{vandehulst45,ewen51,muller51}. This 21-cm radiation can be observed interferometrically at radio wavelengths as a deviation from the brightness temperature of the CMB (\citealt{field58}; \citealt{field59}; \citealt*{madau97}; \citealt{shaver99}). 

Observationally, the 21-cm signal will be accompanied by systematic effects due to the ionosphere and instrument response, system noise, extragalactic foregrounds and Galactic foregrounds \citep[e.g.][]{jelic08,jelic10}, the latter of which are orders of magnitude larger than the 21-cm signal we wish to detect. The foregrounds must be accurately and precisely removed from the observed data as any error at this stage has the ability to strongly affect the EoR 21-cm signal. Foreground removal and the implications for 21-cm cosmology have been extensively researched over the past decade (e.g \citealt{dimatteo02}; \citealt{oh03}; \citealt*{dimatteo04}; \citealt*{zal04}; \citealt{morales04}; \citealt*{santos05}; \citealt{wang06}; \citealt{mcquinn06}; \citealt{jelic08}; \citealt*{gleser08}; \citealt*{bowman06}; \citealt{harker09b}; \citealt*{liu09a}; \citealt{liu09b}; \citealt{harker10}; \citealt{liu11}; \citealt{petrovic11}; \citealt{mao12}; \citealt{liu12}; \citealt*{cho12}; \citealt{chapman12}). This paper concentrates on the spectral fitting of the foregrounds and assumes that bright sources have been accurately removed, for example via a flux cut \citep{dimatteo04}. This is a safe assumption since, as recently shown by \citealt*{trott12}, the residuals from the bright source subtraction are expected to be smaller than the thermal noise contribution and so are not a limiting factor. 

In our first paper, \citet{chapman12}, we introduced a method to successfully remove the foregrounds while making only minimal assumptions. By introducing another similarly successful method, we acknowledge the possibility that different methods will be well suited for the extraction of different information from the data and that there is an advantage in having several foreground cleaning methods to apply the data independently to confirm a statistical detection. In this paper we implement another non-parametric method, the sparsity-based blind source separation (BSS) technique Generalized Morphological Component Analysis, or \textsc{gmca}. \textsc{gmca} provides a complete basis set for foreground removal as opposed to a polynomial fitting method which is not a complete set unless we describe it with a polynomial of the order of the number of frequencies. Polynomial methods rarely utilise orders larger than five and so most polynomial fitting methods may leave the foregrounds incompletely described compared to \textsc{gmca}.

In Section \ref{fgrem} we summarise the various foreground removal pipelines which have been introduced in the literature so far and then go on to detail the method that we utilise, \textsc{gmca}. In Section \ref{sims} we introduce our data cube and the methods used to produce it before presenting the statistical results in Section \ref{results}. In Section \ref{phases} we explore the possibility of recovering images of reionization before we set out our conclusions in Section \ref{conclusions}.

\section{Foreground Removal Techniques}
\label{fgrem}
The statistical detection of the 21-cm reionization signal depends on an accurate and robust method for removing the foregrounds from the total observed signal. For a brief summary of the recent 21-cm foreground removal literature we ask the reader to refer to Section 2 of \citet{chapman12}. 
 
The majority of the literature involves parametric methods, whereby at some point a certain form for the foregrounds is assumed, for example polynomials (e.g. \citealt{santos05}; \citealt{wang06}; \citealt{mcquinn06}; \citealt{bowman06}; \citealt{jelic08}; \citealt{gleser08}; \citealt*{liu09a}; \citealt{liu09b}; \citealt{petrovic11}). In contrast, non-parametric methods do not assume a specific form for the foregrounds, instead allowing the data to determine the foregrounds using more free parameters. While advantageous for poorly constrained data, results are often not as promising as parametric results, though recent methods have challenged this. \citet{harker09b,harker10} preferentially considered foreground models with as few inflection points as possible, which when applied to simulated LOFAR-EoR data compared very favourably with parametric methods. Similarly, \textsc{fastica} as presented in \citet{chapman12} accurately recovered the 21-cm power spectra by considering the statistically independent components of the foregrounds. \textsc{gmca} (\citealt{bobin07}, \citealt{bobin08a}, \citealt*{bobin08b}, \citealt{bobin12}) is another non-parametric method which has a greater flexibility through wavelet choice without the sacrifice of the blind nature of the approach. We will show that \textsc{gmca} not only recovers the power spectra to high accuracy but also that, using wavelet decomposition, the simulated 21-cm signal maps can be recovered exceedingly well after the foreground removal process. 

\subsection{The GMCA method}
The non-parametric method of removing the foregrounds is effectively a BSS problem. \citet{chapman12} utilised a statistical approach to source separation, namely \textsc{fastica}. This assumed that the components of the foregrounds were statistically independent and non-Gaussian in order to reconstruct the smooth spectral form of the foregrounds and leave a residual signal from which we could identify the 21-cm emission statistics. This statistical pursuit of independence is only one form that BSS techniques take, the other utilising morphological diversity and sparsity to separate the sources. \citet{zibulevsky01} proposed a new method of BSS, where one could find a basis set in which the sources to be found would be sparsely represented, i.e. a basis set where only a few of the coefficients would be non-zero. With the sources being unlikely to have the same few non-zero coefficients one could then use this sparsity to more easily separate the mixture. For example, were the 21-cm signal strong enough to detect directly using this method, we would expect it to be sparse on certain scales given the characteristic size of the ionized bubbles, much in the same way the SZ effect can be detected with this method when analysing CMB data \citep{bobin08a}. These bubble sizes change as a function of redshift so the sparse signal from these would arise as a pattern as a function of wavelength. In comparison, the smooth frequency structure of the foregrounds implies that the few sparse non-zero coefficients describing the foregrounds at the same scales as the 21-cm signal would be unchanging with frequency. Given our noise realisations, the 21-cm signal is far too small for this technique to pick it out as a source in its own right. Instead, it is how the foregrounds can be described as different sparse components which enables us to obtain the 21-cm signal and noise as a residual. 

The idea of exploiting the sparseness of sources in different bases has evolved into a full and diverse field of applications. The method has evolved to allow sources to have different morphologies, exploit multichannel data and consider different bases for different sources in order to achieve the most sparse representations. 

Consider an observation of $m$ maps each constituting $t$ pixels across $m$ channels of observation. The problem to be solved can be stated in the following manner:

\begin{equation}
\label{xas}
\mathbfss{X}=\mathbfss{A}\mathbfss{S}+\mathbfss{N}
\end{equation}

\noindent where $\mathbfss{X}$ is the $m \times t$ matrix representing the observed data, $n$ is the number of sources to be estimated, $\mathbfss{S}$ is the signal $n \times t$ matrix to be determined, $\mathbfss{A}$ is the $m \times n$ mixing matrix and $\mathbfss{N}$ is the $m \times t$ noise matrix.

As this is a BSS problem, we need to estimate both $\mathbfss{S}$ and $\mathbfss{A}$. We seek to find the 21-cm signal as a residual in the separation process, therefore $\mathbfss{S}$ represents the foreground signal and, due to the extremely low signal-to-noise of this problem, the 21-cm signal is numerically ignored by the method and can be thought of as an insignificant part of the noise.

We can expand the sources, $\mathbfss{S} = \sum_{j=0}^n \bmath{s}_j$, in a wavelet basis which can mathematically be thought of as a matrix of $T$ wavelet waveforms, $\Phi=[\phi_1,...,\phi_T]$, such that $\forall j \in \{1,...,n\} \quad \bmath{s}_j = \sum_{k=1}^T \alpha_j[k] \phi_k$. $\bmath{s}_j$ is defined to be sparse if only a few of the $\alpha_{j}[k]$ are significantly non-zero. 

The objective of GMCA is to seek an unmixing scheme, through the estimation of $\mathbfss{A}$,
which yields the sparsest sources $\mathbfss{S}$ in the wavelet domain. This is expressed
by the following optimization problem written in the augmented Lagrangian form~:
\begin{equation}
\label{eq:GMCA2}
\min\left(\frac{1}{2}||\mathbfss{X}-\mathbfss{A} \alpha \Phi||_{\mathrm{F}}^{2}+\lambda\sum_{j=1}^{n} ||\alpha_{j}||_{p}\:\right),
\end{equation}
\noindent where typically $\| \alpha \|_p = \left( \sum_k |\alpha[k]|^p\right)^{1/p}$; sparsity is generally enforced for $p=0$ which measures the number of non-zero entries of $\alpha$ (or its relaxed convex version with $p=1$) and $|| \mathbfss{X}||_{\mathrm{F}}=\left(\textrm{trace}(\mathbfss{X}^{T}\mathbfss{X})\right)^{1/2}$
is the Frobenius norm. The problem in Equation~\ref{eq:GMCA2} is solved in an iterative two-step algorithm such that at each iteration $q$~:
\begin{enumerate}
\item{Estimation of the $\mathbfss{S}$ for $\mathbfss{A}$ fixed to ${\mathbfss{A}}^{(q-1)}$~:}\\ \\
Solving the problem in Equation~\ref{eq:GMCA2} for $p=0$ assuming $\mathbfss{A}$ is fixed to ${\mathbfss{A}}^{(q-1)}$, the sources are estimated as follows~:
$$
{\mathbfss{S}}^{(q)} = \Delta_{\lambda}\left({{\mathbfss{A}}^{(q-1)}}^+ {\mathbfss{X}} {\Phi}^T\right){\Phi}
$$
where $\Delta_{\lambda}$ stands for the hard-thresholding operator with threshold $\lambda$; this operator puts to zero all coefficients with amplitudes lower than $\lambda$. In practice, the threshold $\lambda$ is set to be equal to $3$ times the  standard deviation of the noise level to exclude noise coefficients. The term ${{\mathbfss{A}}^{(q-1)}}^+$ denotes the Moore pseudo-inverse of the matrix ${{\mathbfss{A}}^{(q-1)}}$.\\ 
\item{Estimation of the $\mathbfss{A}$ for $\mathbfss{S}$ fixed to ${\mathbfss{S}}^{(q)}$~:}\\ \\
Updating the mixing matrix assuming that the sources are known and fixed to ${\mathbfss{S}}^{(q)}$ is as follows~:
$$
{\mathbfss{A}}^{(q)} = {\mathbfss{X}} {{\mathbfss{S}}^{(q)}}^+
$$
\end{enumerate}
For more technical details about GMCA, we refer the interested reader to (\citealt{bobin07}, \citealt{bobin08a}, \citealt*{bobin08b}, \citealt{bobin12}), where it is shown that sparsity, as used in GMCA, allows for a more precise estimation of the mixing matrix $\mathbfss{A}$ and more robustness to noise than ICA-based techniques.

\textsc{GMCA} provides an efficient method of separating the foreground signal from the noise and 21-cm signal by locating the most sparse components that the foreground signal could be made of in the wavelet basis $\Phi$. From a Bayesian point of view, using this sparsity method is equivalent to having an in-built prior in the model that the foregrounds are sparse over the basis chosen. While a Bayesian evidence analysis could be possible in order to compare different methods as well as different bases, and quantify the overfitting due to more free parameters, we consider it outside the scope of this project.

\subsection{Wavelets}
The set of basis functions, $\Phi$, used by \textsc{gmca} comprises wavelet functions.

The Fourier transform is a well known method of analysing data at different scales with a single set of basis functions - sines and cosines. In reality this confined basis set can obscure information, and so we instead consider an infinite set of basis functions localized in space - the wavelet functions. There are many types of wavelets - with some more localized in space, some smoother and some with fractal structures.

The most common form of wavelet used in astrophysics is the isotropic undecimated wavelet transform (IUWT) which we describe briefly below in reference to the more complete descriptions in literature (e.g. \citealt*{starck98}, \citealt{starck06}). Consider an image with $p \times p$ pixels (where, using the previous section's notation $p \times p = t$ and we will refer to the pixel coordinates as $[k,l]$). 

We can decompose an image at a particular frequency, $c_0$, into a coarse version of itself, $c_J$, along with a superposition of the original image at different wavelet scales:

\begin{equation}
\label{IUWT}
 c_0[k,l] = c_J[k,l]+\sum_{j=1}^{J}w_j[k,l]
\end{equation}

\noindent where the wavelet coefficient $w_j$ represents the data at scale 2$^{-j}$.

The decomposition is typically achieved using low-pass 1D filters, which we call $h_{1D}$, implemented by the ``\`{a} trous'' algorithm:

\begin{eqnarray}
\label{atrous}
&& c_{j+1}[k,l]=\sum_q\sum_p h_{1D}[p]h_{1D}[q]c_j[k+2^jq,l+2^jp] \\
&& w_{j+1} = c_j[k,l] - c_{j+1}[k,l]
\end{eqnarray}

\noindent When $c_0$ can be described by only a few significantly non-zero $w_j$, we say that $c_0$ is sparse in that basis. 

In this paper we utilise wavelets twice. The first time is within the \textsc{gmca} algorithm, where we have a choice of different wavelet types that can be used as the basis set. \textsc{gmca} uses wavelet decomposition to identify the sources, $\mathbfss{S}$, but then returns a data cube with all scales present. In our analysis in Section \ref{phases} we wish to look at these results on different scales and so we utilise wavelet decomposition to do this. We will use the IUWT both within \textsc{gmca} and later to analyse the images at different scales, though we briefly consider other wavelets when we question how much our results depend on this choice of wavelet.

For our \textsc{gmca} implementation we set the number of decomposition scales to be eight, the maximum allowed by the dimensions of our data cube.

\section{Simulated EoR Data}
\label{sims}
We simulate 170 frequency maps between 115 and 199.5 MHz with spacings of 0.5 MHz. The maps consist of $512^2$ pixels representing a comoving 1734 Mpc$^2$ area. At $z=$10.8 this is equivalent to a field of view of 10$^\circ \times$10$^\circ$ or a resolution of 1.17 arcminutes per pixel. Since an interferometer like LOFAR is insensitive to the mean value of the brightness temperature, we use mean-subtracted maps. We use the same set of simulations as \citet{chapman12} and the reader can refer there for more detailed information on the simulations. 

The observable of the 21-cm radiation, the brightness temperature $\delta T_\mathrm{b}$, is simulated using the semi-numeric modelling tool \textsc{21cmfast} (\citealt{mesinger07}; \citealt*{mesinger11}). The code was run using a standard cosmology, ($\Omega_{\Lambda},\Omega_m,\Omega_b,n,\sigma_8,h$)=(0.72,0.28,0.046,0.96,0.82,73)\citep{komatsu11} and initialised at $z$=300 on a $1800^3$ grid. The velocity fields used to perturb the initial conditions as well as the resulting 21-cm $\delta T_\mathrm{b}$ boxes were formed on a cruder grid of $450^3$ before being interpolated up to $512^3$. We define halos contributing ionizing photons as having a minimum virial mass of $1\times 10^9 \mathrm{M\odot}$. 

The foreground simulations are obtained using the foreground models described in \citet{jelic08,jelic10}. We model Galactic diffuse synchrotron emission (GDSE), Galactic localised synchrotron emission, Galactic diffuse free-free emission and extragalactic foregrounds consisting of contributions from radio galaxies and radio clusters. While the Galactic foreground emission is simulated using Gaussian random fields, the total foreground signal is non-Gaussian. We do not consider the polarisation of the foregrounds. The foregrounds simulated here can be up to five orders of magnitude larger than the signal we hope to detect but since interferometers such as LOFAR measure only fluctuations, foreground fluctuations dominate by `only' three orders of magnitude. 

To simulate the noise at each frequency, a LOFAR measurement set was filled with Gaussian noise in the uv plane. This was then imaged to create a real-space image, the root mean square of which can be normalized to the value as given by the prescription detailed in \citet{chapman12}. For example the noise sensitivity at 150 MHz for an integration time of 600 h, a PSF of width 4 arcminutes and a frequency spacing of 0.5 MHz is 64 mK. The 170 noise maps were uncorrelated over frequency - i.e. a different noise realization was used to fill the measurement set for each frequency. 

The success of an interferometer such as LOFAR is highly dependent on how uv space is sampled. The particular pattern of uv sampling forms a beam which affects how the components such as the foregrounds are seen by the interferometer. Dirty foreground and 21-cm images were simulated by convolving with the PSF of the LOFAR set-up used to simulate the noise. The PSF used for creating dirty images (and for creating the noise as described in the previous section) was chosen to be the worst in the observation bandwidth - i.e. the PSF at 115 MHz. In observations the synthesized beam decreases in size with increasing frequency, causing point source signals to oscillate with the beam, producing a foreground signal with an oscillatory signal very much like that of the 21-cm signal \citep*{vedantham12}. However, this mode-mixing contribution has been found not to threaten the 21-cm recovery and have a power well below the 21-cm level (\citealt{bowman06}; \citealt*{liu09a}; \citealt*{trott12}). As such we do not consider a frequency dependent PSF here.

\begin{figure}
\includegraphics[width=84mm]{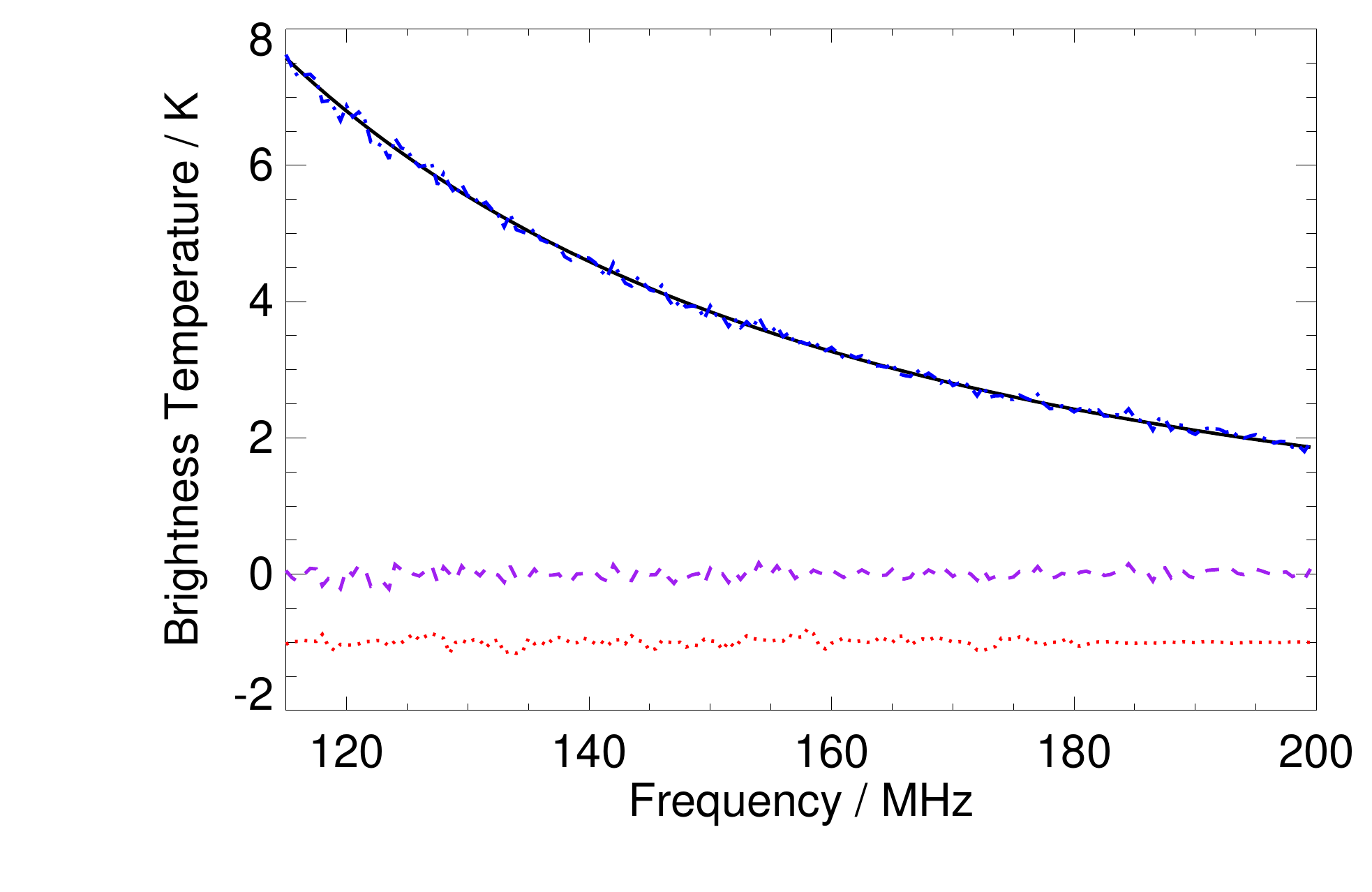}
\caption{The redshift evolution of the simulated cosmological signal (red; dot), foregrounds (black;solid), noise (purple; dash) and total combined signal (blue; dash dot). All components have undergone the PSF convolution. Note the 21-cm signal has been amplified by 10 and displaced by -1K for clarity.}
\label{sims_los}
\end{figure}

Once the foregrounds and 21-cm signal have been adjusted for uv sampling, the three component cubes are added together. The components of the total $\delta T_\mathrm{b}$ along a random line of sight are shown in Fig. \ref{sims_los}.

\section{Results}
\label{results}
In the following section, the word `reconstructed' refers to a component which has been estimated from the simulated data using \textsc{gmca}. The `residuals' or (`rest' for short) are the difference between the total mixed signal and the reconstructed foregrounds and should therefore consist of the 21-cm signal, noise and any fitting errors.
\subsection{Source Number}
\label{source}
\textsc{gmca} requires the specification of the number of sparse sources that the data can be defined by. We utilise this component separation technique to define the foregrounds in order to subtract them, treating the 21-cm signal as noise. Therefore, the number of sources refers to the number of foreground contributions which can be described by unique sparse descriptions (not necessarily the number of different foreground components such as Galactic free-free). 

As each foreground model might be best described by a different number of sources, we seek to define the number of sources by minimizing the leakage of foregrounds into the 21-cm signal. Let us introduce the statistics $\mathbfss{L}_Y$ and $\mathbfss{R}_Y$:

\begin{equation}
\mathbfss{L}_Y = \mathbfss{A} (\mathbfss{A}^T \mathbfss{A})^{-1} \mathbfss{A}^T \mathbfss{Y} 
\label{leakeqn}
\end{equation}

\begin{equation}
\mathbfss{R}_Y = \mathbfss{Y} - \mathbfss{A} (\mathbfss{A}^T \mathbfss{A})^{-1} \mathbfss{A}^T \mathbfss{Y} 
\end{equation}

\noindent where $\mathbfss{A}$ is the mixing matrix calculated by \textsc{gmca} and $\mathbfss{Y}$ is a data cube, for example the foregrounds or the 21-cm signal. $\mathbfss{L}_Y$ is the amount of the data $\mathbfss{Y}$ that contributes to the \textsc{gmca} sources (in our case the reconstructed foregrounds). Thus we can calculate the amount of leakage of simulated noise and 21-cm signal, $\mathbfss{L}_{nocs}$ into the reconstructed foregrounds by allowing $\mathbfss{Y}$ to equal the combined simulated 21-cm and noise data cube. Conversely, $\mathbfss{R}_Y$ is the amount of data $\mathbfss{Y}$ which does not contribute to the \textsc{gmca} source model. For example, letting $\mathbfss{Y}$ equal the simulated foreground cube, $\mathbfss{R}_{fg}$, will tell us how much of the foregrounds leak into the residuals.

\begin{figure}
\includegraphics[width=84mm]{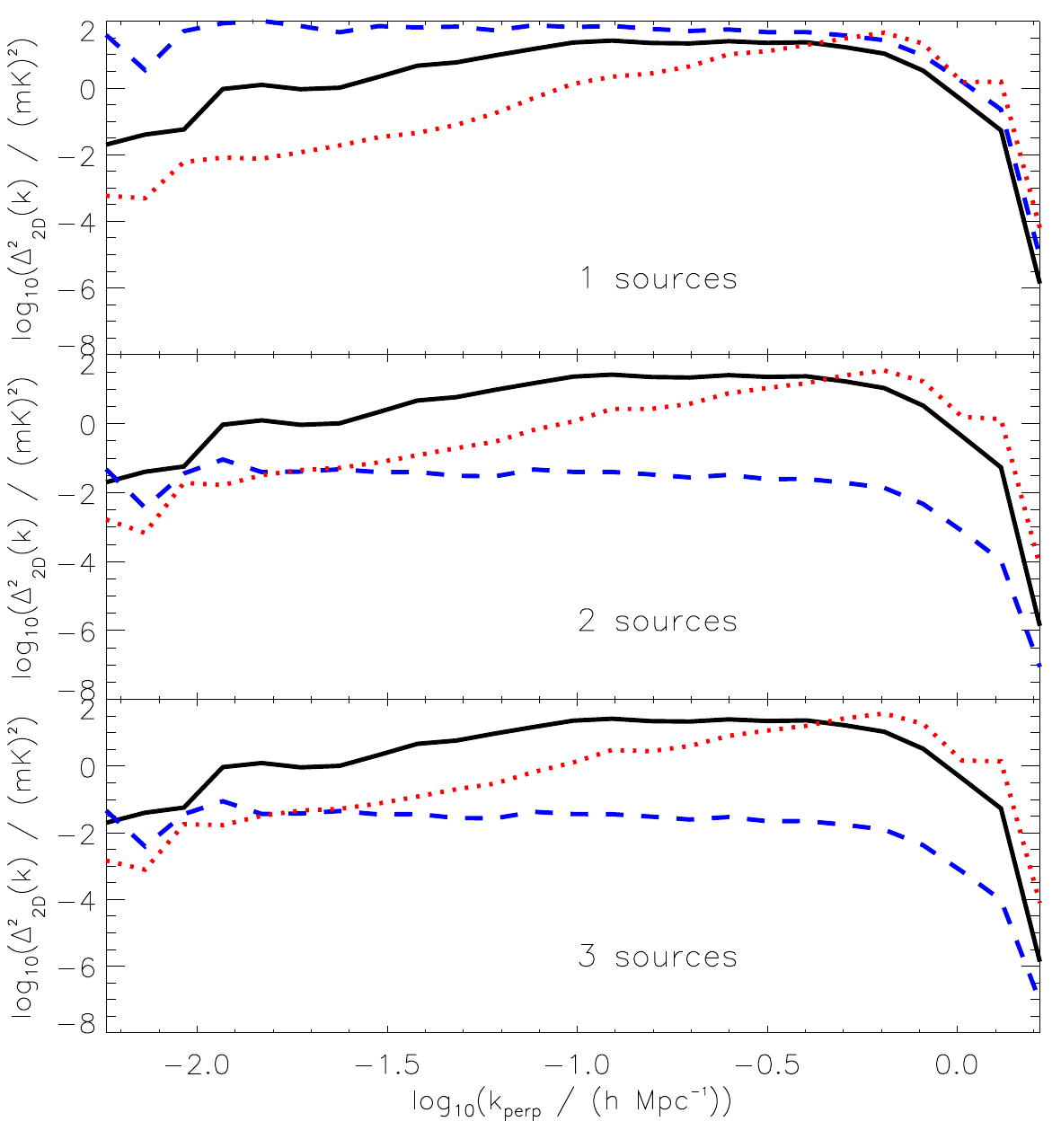}
\caption{The 2D power spectrum of $\mathbfss{R}_{fg}$ (blue dash) and $\mathbfss{L}_{nocs}$ (red dot) and the 21-cm power spectrum (black,solid) at a frequency of 160.0 MHz for \textsc{gmca} with 1, 2 and 3 sources (top, middle and bottom respectively).}
\label{so_no_leak}
\end{figure}

We take the power spectra of $\mathbfss{L}_{nocs}$ and $\mathbfss{R}_{fg}$ and compare them to the power spectra of the 21-cm signal to see the number of sources for which leakage is minimized, Fig. \ref{so_no_leak}. Note that, prior to our wavelet type analysis in Section \ref{wchoice_sec}, we adopt the default setting of \textsc{gmca} for this source number analysis, the IUWT using the \`{a} trous algorithm.
 
To be confident of our reconstructed 21-cm signal, we ideally want both the power spectra of $\mathbfss{R}_{fg}$ and $\mathbfss{L}_{nocs}$ to lie below that of the 21-cm signal and to be as small as possible. We see that, while one source does not seem enough to accurately constrain the foregrounds, there is very little difference between the leakages resulting from a  2 or 3 source foreground model. Indeed, we find this holds true for 4 and 5 sources also. As more and more sources are added to the model, we might expect the 21-cm signal itself to leak into the foreground model and be picked out as an individual source. However the magnitude of this (seen as part of $\mathbfss{L}_{nocs}$) will be small as the signal to noise of the foregrounds is much larger than that of the cosmological signal. When applying to real data we will have to rely on models of the foregrounds as observed by the data to estimate the level of leakage from the signal onto the foregrounds.

We choose to assume two foreground sources for the rest of this paper, though the reader should be aware that different foreground models may require different source numbers in order to optimise the method. 

\subsection{Choice of wavelet}
\label{wchoice_sec}
\begin{figure}
\includegraphics[width=84mm]{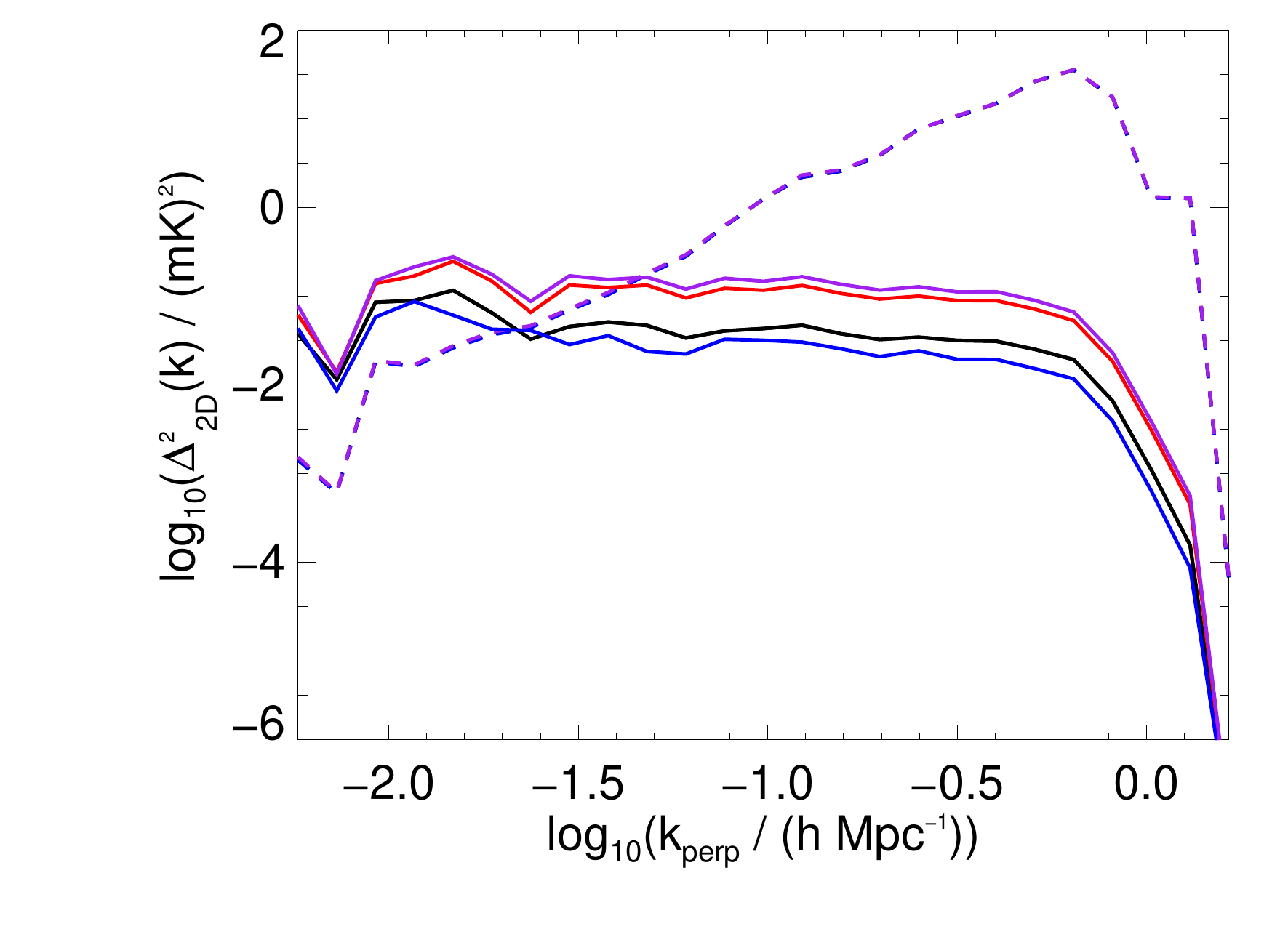}
\caption{The 2D power spectrum of $\mathbfss{R}_{fg}$ (solid) and $\mathbfss{L}_{nocs}$ (dashed) at 160 MHz for: IUWT (black), Mallat's wavelet transform (red), Feauveau's wavelet transform without under-sampling (blue), Haar's wavelet transform (purple). Note that the dashed lines all lie on top of each other.}
\label{wchoice}
\end{figure}

In Fig. \ref{wchoice} we consider how the leakages $\mathbfss{R}_{fg}$ and $\mathbfss{L}_{nocs}$ change depending on the choice of wavelet used by \textsc{gmca}. There are many different types of wavelets, but they can be broadly categorized by whether they are decimated (i.e. provide a redundant signal representation), isotropic and which filter they use for the separation of data at different scales. Here we consider the IUWT, a decimated wavelet transform (referred to as Mallat), a non-dyadic and undecimated wavelet transform (referred to as Feauveau) and an undecimated wavelet transform using the Haar filter (referred to as Haar).

We can see that the choice of wavelet can affect how small the foreground leakages are, and therefore the success of the method. These differences are highly dependent on frequency as well, with one wavelet type out-performing others at certain frequencies. For our implementation we choose the IUWT as it consistently minimizes the leakages over the frequency range. We will show in Section \ref{lps} that we can potentially correct for $\mathbfss{L}_{no}$, though not $\mathbfss{L}_{cs}$ and $\mathbfss{R}_{fg}$ due to the blind nature of the problem.

\subsection{Power Spectra}
\label{ps}
EoR experiments aim to recover the power spectrum of the cosmological signal over a broad range of frequencies.

The power spectrum of a line-of-sight/map/cube at a single frequency is calculated by 1D/2D/3D Fourier transforming that line-of-sight/map/cube and binning the pixels according to Fourier scale, $k$. The power at any particular $k$, $\langle\delta(k) \delta^*(k)\rangle$ is the average power of all the uv cells in the bin centering on $k$. The error on the point for a particular bin, $k_i$, is calculated as $\sigma_i = \frac{\langle\delta(k_i) \delta^*(k_i)\rangle}{\sqrt{n_{k_i}}}$ where $n_{k_i}$ is the number of uv cells that reside in that $k$ bin. The power spectrum of the reconstructed 21-cm signal is calculated by subtraction of the noise power spectrum from the \textsc{gmca} residuals power spectrum. The total error on the reconstructed 21-cm power spectrum is calculated using the above error formula applied to the reconstructed 21-cm power spectrum, added in quadrature with the formula applied to the noise power spectrum, in order to take into account both sample variance and the effect of any error in the noise estimate. Note that we assume Gaussianity whereas the 21-cm signal is not Gaussian and also we calculate the error bars from the power of a single realization rather than over an ensemble of simulations. We ask the reader to bear in mind that these error bars might be considered incomplete because of this. 

To explain a few graphical conventions: any points where the power of the residuals is below the power of the noise are omitted, as this leads to an unrealistic negative reconstructed 21-cm power; any error bars extending to below the x axis in linear space are shown with a lower error bar of equal length to the upper error bar in log space. 

\subsubsection{1D Power Spectra}
\begin{figure}
\includegraphics[width=84mm]{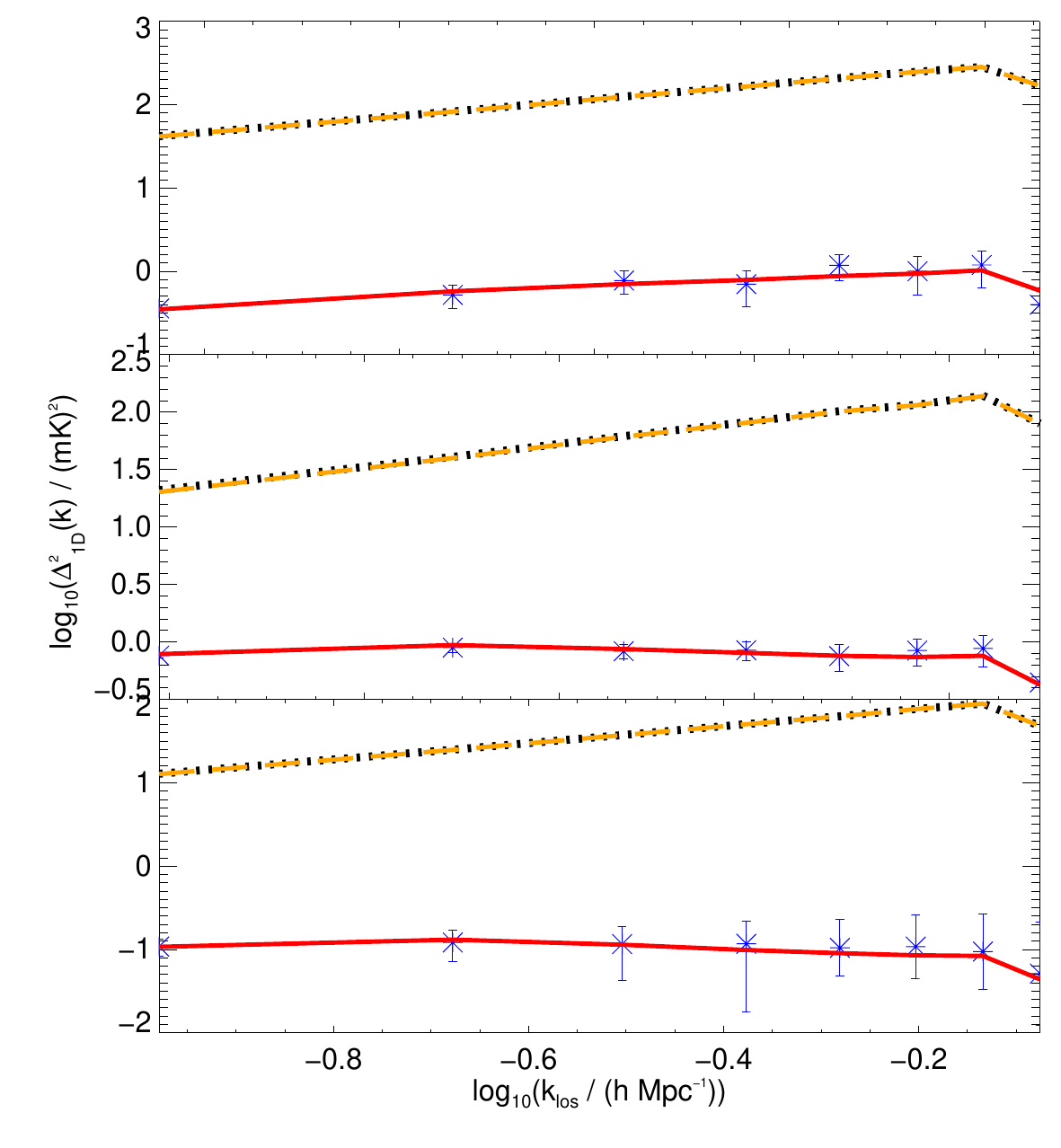}
\caption{The 1D power spectra of the simulated 21-cm (red, solid), the residuals (black, dot), the reconstructed 21-cm (blue, points) and the noise (orange, long dash). Three 8 MHz frequency wedges centred at 127 MHz, 151 MHz and 175 MHz respectively are shown from top to bottom.}
\label{1Dps}
\end{figure}

The 1D power spectra are calculated over frequency wedges of 8MHz to avoid evolution effects. Each line of sight produces a 1D power spectrum, one for each of the 512 $\times$ 512 pixels. These power spectra are then averaged over all the pixels.  The frequencies mentioned correspond to the frequency in the middle of each 8MHz wedge and the quantity plotted in Fig. \ref{1Dps} is $\Delta^2_{1D}(k) = \frac{L k \langle\delta(k) \delta^*(k)\rangle}{\pi}$ where $L$ is the comoving length of the wedge. The 1D power spectra are recovered to high accuracy across the frequency range and across the scales.

\subsubsection{2D and 3D Power Spectra}
\label{lps}
For the 2D power spectra, the quantity plotted is $\Delta^2_{2D}(k) = \frac{A k^2 \langle\delta(k) \delta^*(k)\rangle}{2 \pi}$ where $A$ is the area of the simulation map. To calculate the 3D power spectra we divide the cube into sub-bands of 8 MHz to avoid signal evolution effects. The quantity plotted is $\Delta^2_{\mathrm{3D}}(k) = \frac{V k^3 \langle\delta(k) \delta^*(k)\rangle}{2 \pi^2}$ where V is the volume of the sub-band.

We now consider the recovered 2D and 3D 21-cm power spectra and how best we can minimise the effects of leakage. The noise leakage into the reconstructed foregrounds can be accurately quantified by creating an independent realization of the noise (no2) (for real data it is assumed we will know the statistics of the noise to high precision and can therefore create a second realization from the known noise power spectrum) and applying Equation \ref{leakeqn} to find $\mathbfss{L}_{no2}$. We find that the power spectra of  $\mathbfss{L}_{no}$ and $\mathbfss{L}_{no2}$ are almost identical meaning that $\mathbfss{A}$ is not a strong function of the original noise realization. We can use this information to find a better estimate of the 2D and 3D power spectra and compensate for the noise leakage using the following calculation:

\begin{equation}
\tilde{\Delta}^2_{csrec} = \Delta^2_{rest} - \Delta^2_{no} + \Delta^2_{L_{no2}}
\label{no2comp}
\end{equation}

\begin{figure}
\includegraphics[width=84mm]{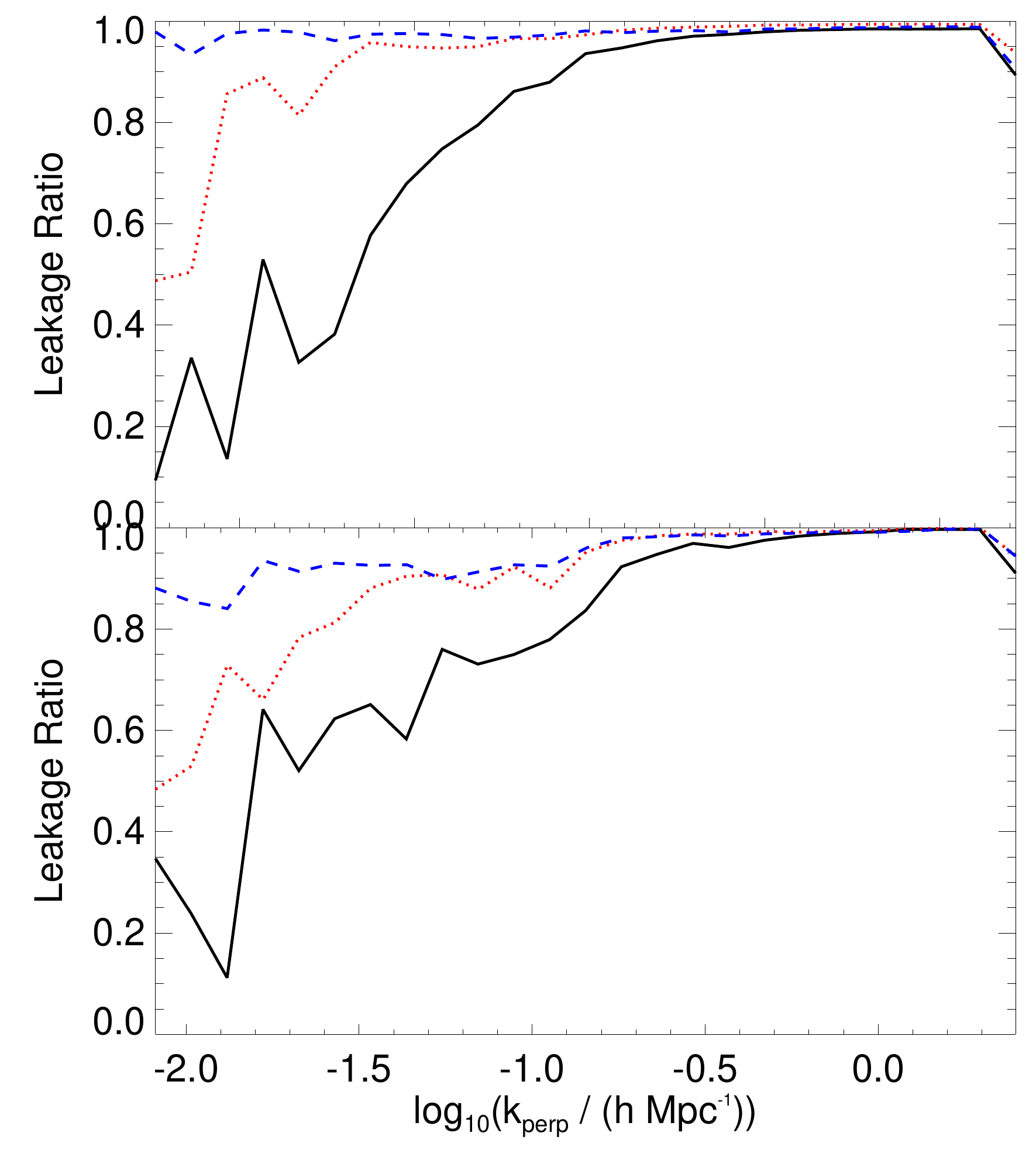}
\caption{Top : The leakage ratio for the 2D power spectra for frequencies of 130MHz (black,solid), 150MHz (red,dot) and 170MHz (blue,dash). Bottom : The leakage ratio for the 3D power spectra for frequencies of 135MHz (black,solid), 151MHz (red,dot) and 167MHz (blue,dash).}
\label{leakage}
\end{figure}

To calculate a leakage ratio we would generally have to include all the leakages, such that $\mbox{Leakage Ratio} = \frac{\Delta^2_{cs}}{\Delta^2_{R_{fg}}+\Delta^2_{L_{nocs}}+\Delta^2_{cs}}$. However, since we can correct for $\mathbfss{L}_{no}$ we can instead calculate:

\begin{equation}
\mbox{Leakage Ratio} = \frac{\Delta^2_{cs}}{\Delta^2_{R_{fg}}+\Delta^2_{L_{cs}}+\Delta^2_{cs}}
\end{equation}

\noindent which quantifies the amount of 21-cm leakage into the foregrounds and vice versa. We can see the leakage ratio plotted for the 2D and 3D power spectra for multiple frequencies in Fig. \ref{leakage}. We see that we can expect an accurate 21-cm power spectrum recovery at large $k$ once the noise leakage is compensated for. The smaller $k$ recovery is still very dependent on the foreground magnitude and therefore frequency of observation. 

\begin{figure*}
\includegraphics[width=110mm]{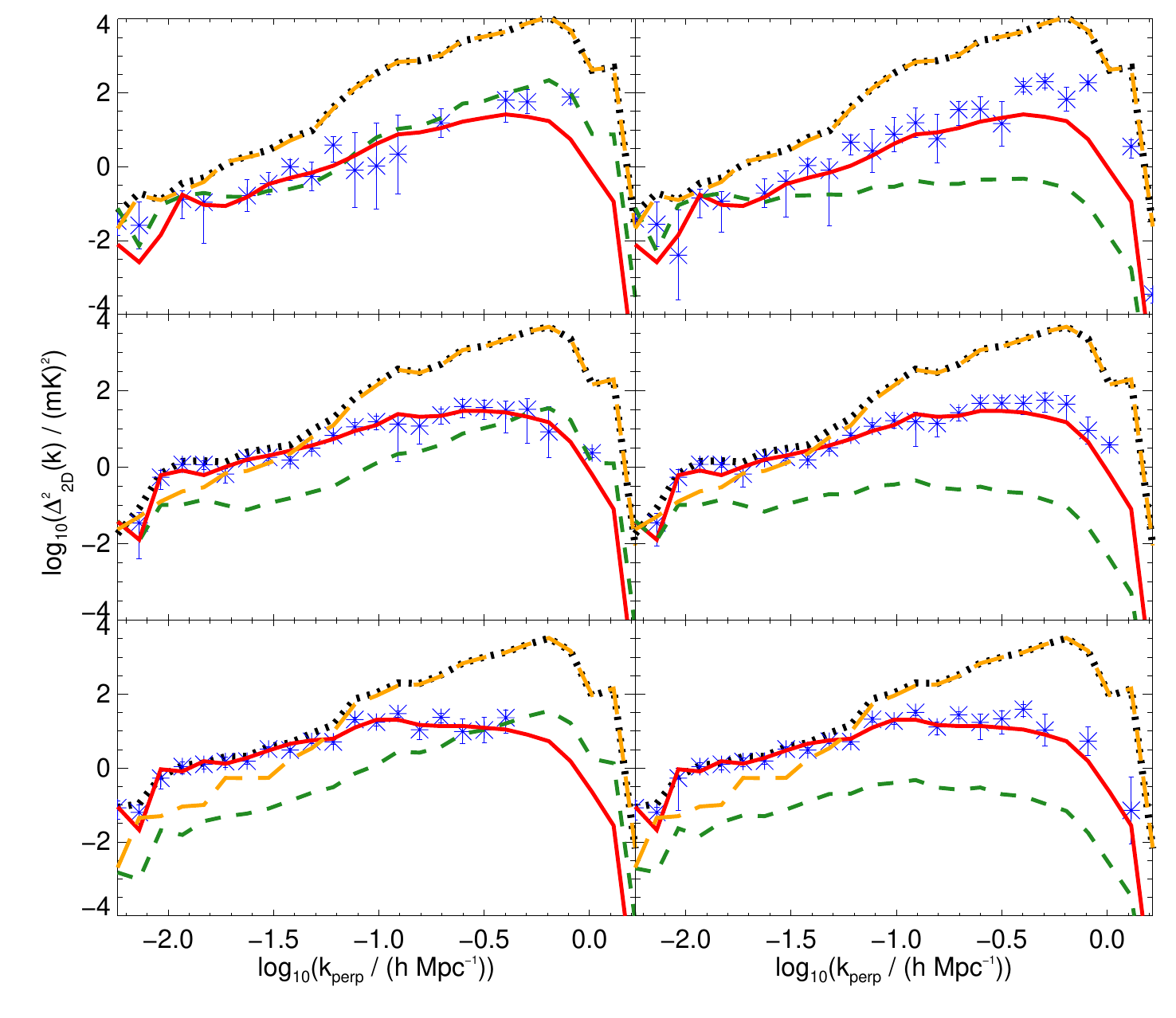}
\caption{2D power spectrum of the simulated 21-cm signal, reconstructed 21-cm signal, residuals and noise at 130 MHz, or $z$=9.92, 150 MHz, or $z$=8.47 and 170 MHz, or $z$=7.35 from top to bottom. The left column is the fiducial data whereas the right hand column plots the reconstructed 21-cm power spectrum but with the leakage determined from the second noise realization added, as described in Section \ref{lps}. Linestyles are as described in Fig. \ref{1Dps} with the additional dark green dashed line representing the total leakage power ($\Delta^2_{R_{fg}} + \Delta^2_{L_{nocs}}$) in the left column and the leakage assuming noise leakage has been corrected ($\Delta^2_{R_{fg}} + \Delta^2_{L_{cs}}$) in the right column.}
\label{2D_ps}
\end{figure*}

\begin{figure*}
\includegraphics[width=110mm]{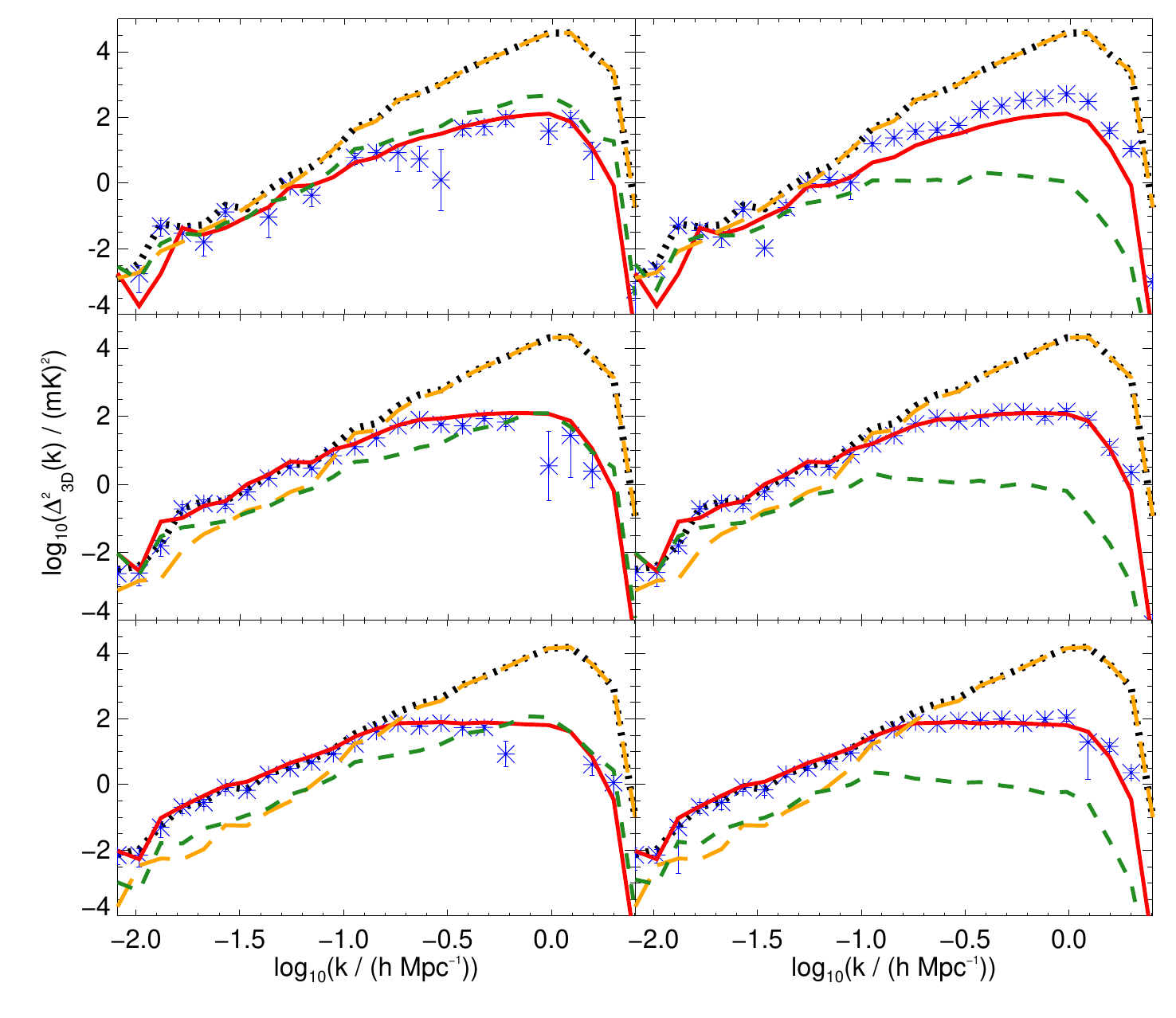}
\caption{3D power spectrum of the simulated 21-cm signal, reconstructed 21-cm signal, residuals and noise at 135 MHz, or z=9.51, 151 MHz, or z=8.40 and 167 MHz, or z=7.50 over an 8 MHz sub band (top to bottom). The left column is the fiducial data whereas the right hand column plots the reconstructed 21-cm power spectrum but with the second noise realization leakage added. Linestyles are as described in Fig. \ref{2D_ps}.}
\label{3D_ps}
\end{figure*}

We can see the result of applying Equation \ref{no2comp} in Figs. \ref{2D_ps} and \ref{3D_ps}. Wherever $\mathbfss{R}_{fg}$ and/or $\mathbfss{L}_{nocs}$ exceeds the power of the simulated 21-cm signal we see a degraded fit in Fig. \ref{2D_ps}. The 2D power spectra are recovered to excellent accuracy and we see that once the noise leakage is taken into account there is much less leakage at large $k$ scales, allowing a more complete power spectrum reconstruction. 

For the 3D power spectra, a similarly accurate recovery is made across the frequency range. The recovery in 3D is more precise due to the larger amount of data in a box as opposed to a single slice. Again, once the noise leakage is taken into account the recovered spectra become much more complete on the larger $k$ scales. 

\section{Phase Conservation and Imaging}
\label{phases}
Recently it has been shown that imaging of the neutral hydrogen in the late stages of reionization is possible with the current generation of radio telescopes when angular scales larger than 0.5$^\circ$ are considered, independent of the type of reionization source \citep{zaroubi12}. Here, we compare the output residual maps with the simulated 21-cm maps and consider how well the phases of the 21-cm signal are conserved through the foreground removal process. The better the phases are conserved, the more correlation between maps we will observe. We will also consider the maps at different scales and as such we also decompose the output maps into 8 wavelet scales using the IUWT.

For a particular frequency, we calculate the phase of each uv point in a Fourier transformed map, F, as Phase[u,v] = $\tan^{-1}$(Im(F(u,v))/Re(F(u,v))). 

\begin{figure*}
\hspace*{2.5cm}
\includegraphics[width=215mm]{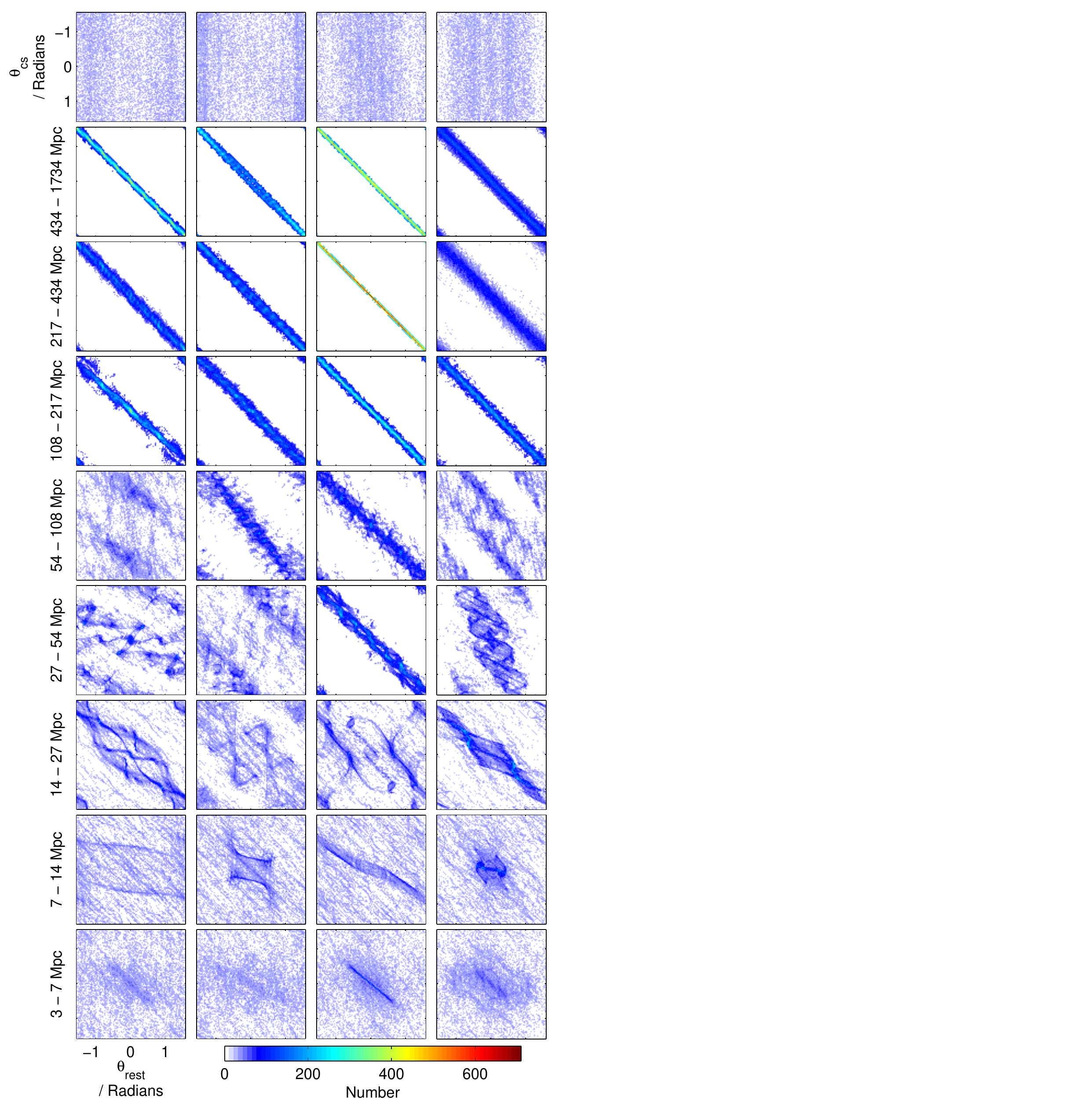}
\caption{Density maps of the phase of maps of the simulated 21-cm and the residuals. From left to right are maps at frequencies 130, 145, 160 and 175 MHz. From top to bottom are maps of the complete cubes and then of increasingly small scale wavelet scales. A clear diagonal signifies excellent phase recovery and therefore clearer images can be recovered. We see that on scales above 108 Mpc, the phases are well preserved; on smaller scales however, the phases are highly uncorrelated. It is clear that considering different wavelet scales can result in much better phase recovery than considering the full cube.}
\label{ph_map}
\end{figure*}

In Fig. \ref{ph_map} we see the phase density relating to each wavelet scale of the 21-cm and residual cubes. For each frequency we calculate the phase of each pixel in the 21-cm and residual maps and use this as a coordinate on a phase density map where the x axis is the phase of the residuals and the y axis is the phase of the 21-cm map. The more pixels with coordinates corresponding to a particular bin in the phases, the higher the phase density we will observe. If \textsc{gmca} perfectly preserves the phase of the 21-cm signal we should see a diagonal phase density plot. For clarity, only the phase bins with a pixel count in the largest 67$\%$ of the pixel count distribution are plotted. We see that for the three crudest wavelet scales there is excellent phase recovery across the frequency range, while at 160 MHz this excellent recovery can be seen on even smaller wavelet scales. This is in line with expectations since the relative rms values (i.e. ratio of the rms of the 21-cm and noise and ratio of the rms of the 21-cm and foregrounds) for the 21-cm signal both peak at just after 160 MHz.

\begin{figure*}
\hspace*{1.5cm}
\includegraphics[width=210mm]{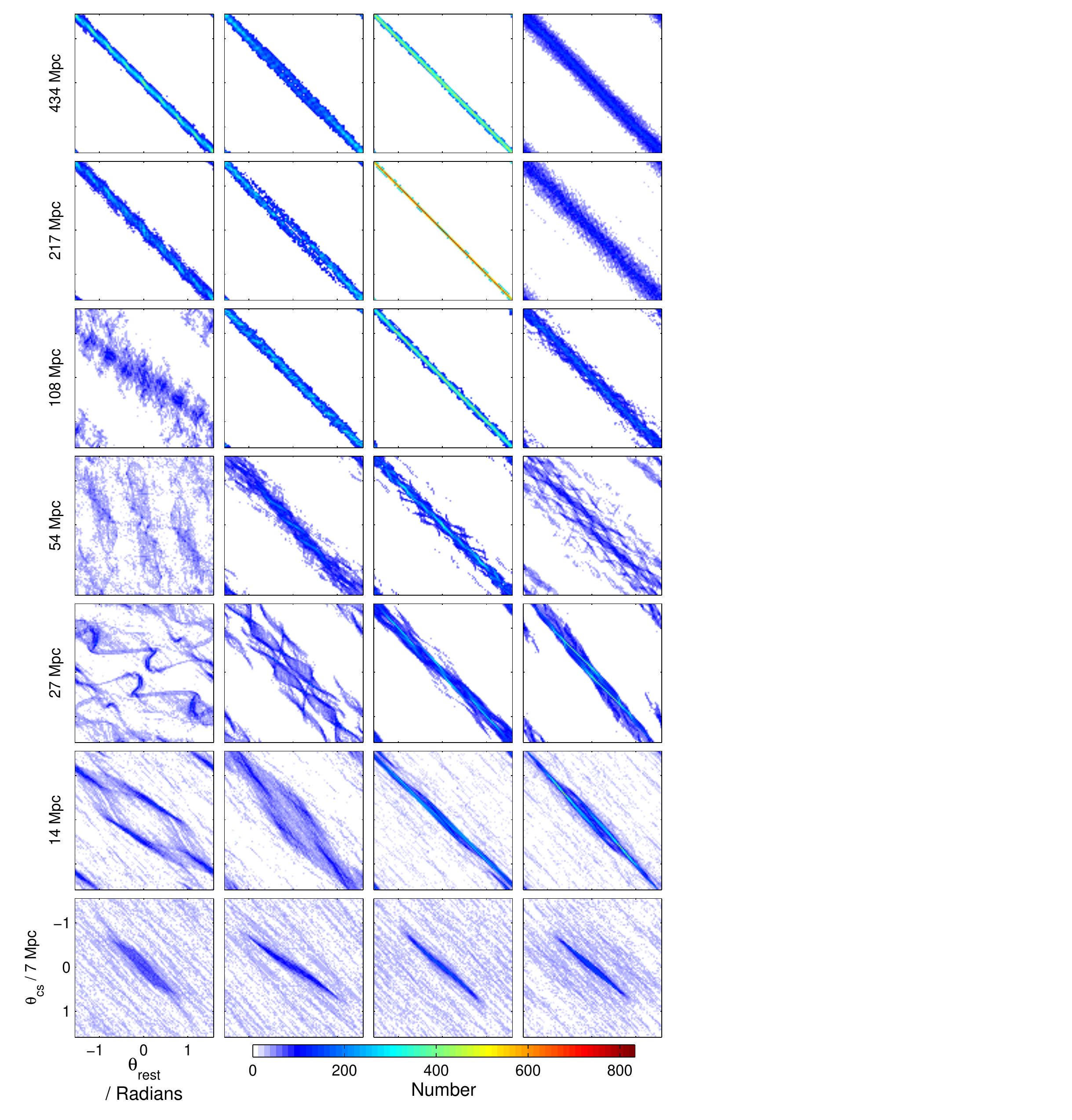}
\caption{Density maps of the phase of maps of the simulated 21-cm and the residuals. From left to right are maps at frequencies 130, 145, 160 and 175 MHz. From top to bottom are maps of cubes with only the crudest scale present and then of only the 2,3,4,5,6 and 7 crudest wavelet scales. The minimum distance scale information included is labelled for each wavelet scale, the maximum is always 1734 Mpc. A clear diagonal signifies excellent phase recovery and therefore clearer images can be recovered. The addition of several scales together results in clearer diagonals than considering scales individually in Fig. \ref{ph_map}.}
\label{ph_map_2}
\end{figure*}

In Fig. \ref{ph_map_2} we start with the crudest wavelet scale and then add the next crudest scale one at a time to see the effect on phase conservation, finding that this improves the recovery considerably.

\begin{figure*}
\includegraphics[width=205mm]{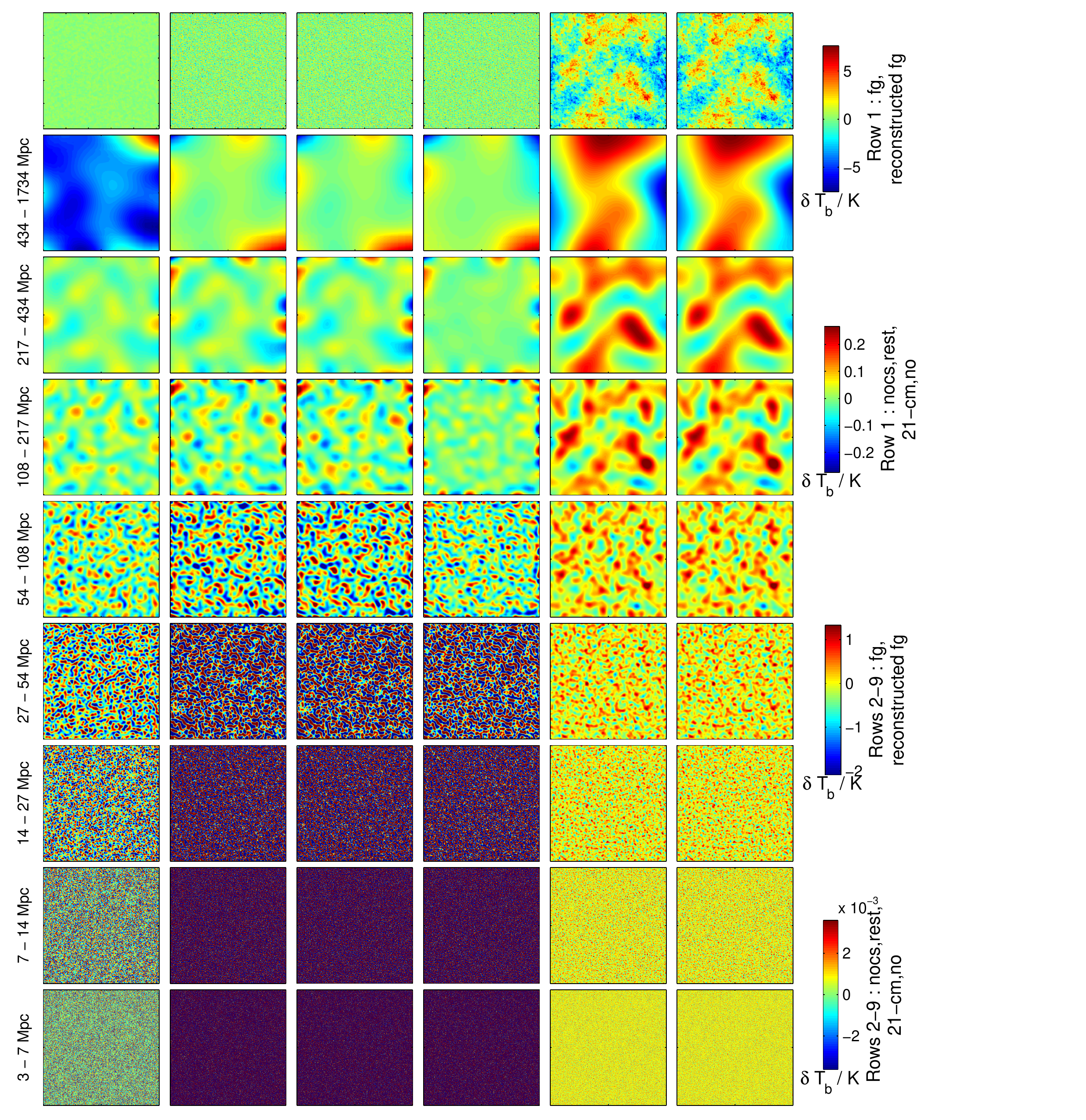}
\caption{The decomposition of the 21-cm signal, residuals,simulated noise + 21-cm signal, noise, foregrounds, reconstructed foregrounds and noise (left to right). From top to bottom, the rows are the original image at 165MHz, and then the wavelet decomposition of this image at the 8 wavelet scales. We can see that the simulated and reconstructed foregrounds have a high correlation at all scales and even in the full cube. Similarly, the noise + 21-cm and residuals also share this strong correlation. As we cannot remove the noise directly we must look for a correlation between the residuals and simulated 21-cm, which will come as a result of little or no correlation between the noise and residuals at certain scales. The noise dominates too much in the full cube and on the large k scales, however we can clearly see a correlation by eye on distance scales between 108 and 434 Mpc. At the largest scale, the 21-cm signal is so small that the residuals are dominated by noise.}
\label{csfgno_decomp}
\end{figure*}

In Fig. \ref{csfgno_decomp} we compare the reconstructed signal maps with the simulated signal maps at the different wavelet scales. It is clear that the foregrounds are reconstructed to a high accuracy at all scales. The residuals show clear correlation with the 21-cm maps - especially at scales 54-434 Mpc. By considering the data on different scales we compensate for the small scale noise leakage and can retrieve convincing reconstructed images in comparison to the map with all scale data included.  

\begin{figure}
\includegraphics[width=84mm]{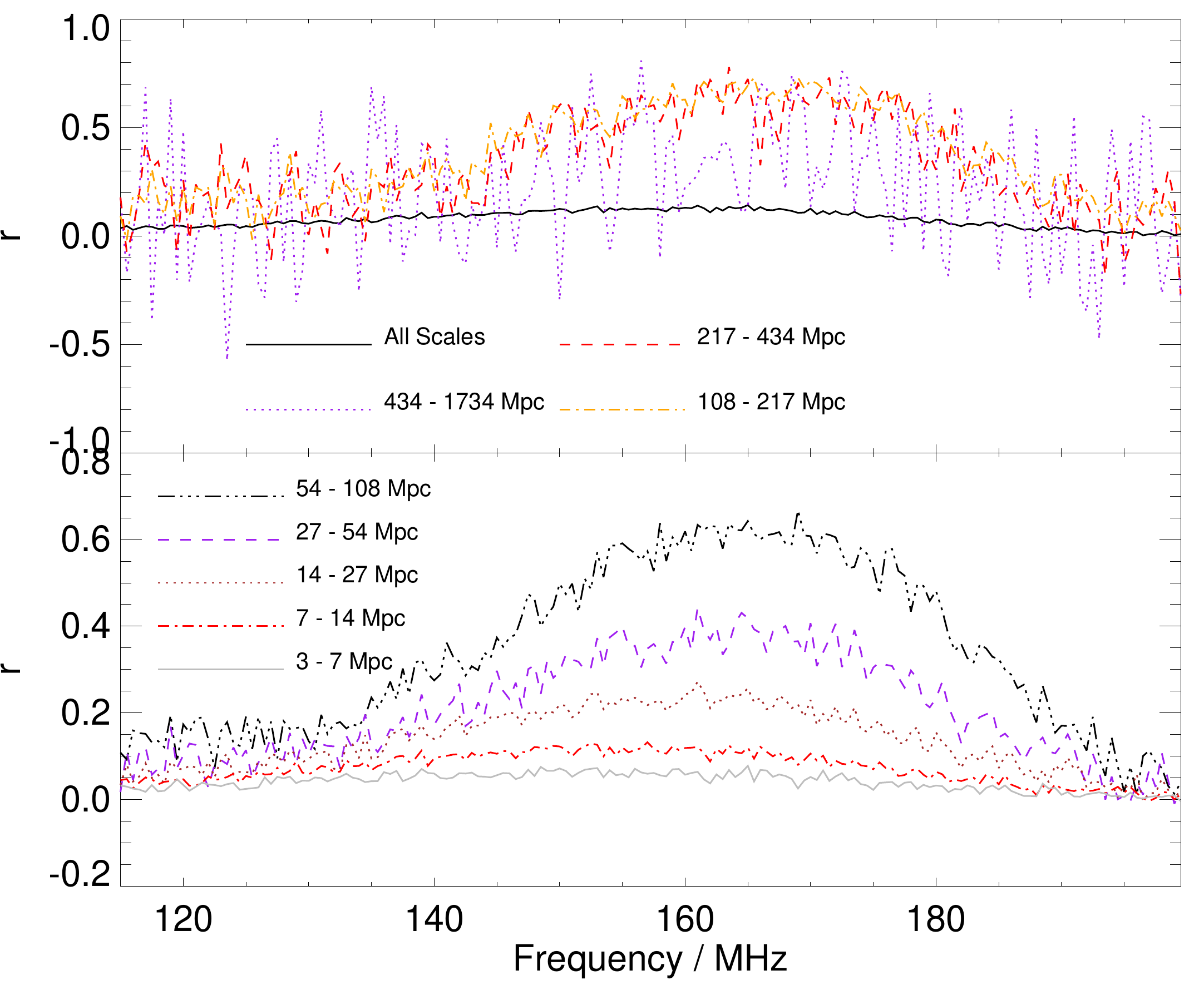}
\caption{The Pearson correlation coefficient between 21-cm and residual maps. There is very little correlation when all scales are present or on the smallest scales, however we reach correlation coefficients of over 0.6 for distance scales 54 - 434 Mpc. The correlations are always much weaker at the lower end of the frequency range because the noise and foregrounds are at their highest and at the higher end of the frequency range because the 21-cm signal is negligible.}
\label{pearson_maps}
\end{figure}

We plot the Pearson correlation coefficient between the simulated 21-cm maps and the residual maps at different individual scales in Fig. \ref{pearson_maps}. On an individual basis, we can clearly see that the finer the scale of structure we correlate, the less of a correlation there is between the residuals and simulated 21-cm maps. The finer wavelet scale we look at, the more dominant noise leakage will be in the residual maps and so a smaller correlation is observed. By only comparing the crudest wavelet scales however, we risk losing a lot of the small scale 21-cm structure and being increasingly dominated by the foreground signal.

We can add several of the wavelet scales together in order to balance having enough useful information without including too much noise leakage on the smaller wavelet scales. To compare with \citet{zaroubi12}, we also include images of the full data which have been smoothed with a 20 arcminute Gaussian kernel. Wavelet decomposition has the advantage of providing a selection of scales on which one can analyse the images, as opposed to being restricted by filters such as a Gaussian kernel which simply remove all modes below a certain scale. However, the scales at which one can analyse images with wavelet decomposition are determined by the method itself - you cannot then ask what the data look like at a scale half way inbetween two wavelet scales.

\begin{figure}
\includegraphics[width=100mm]{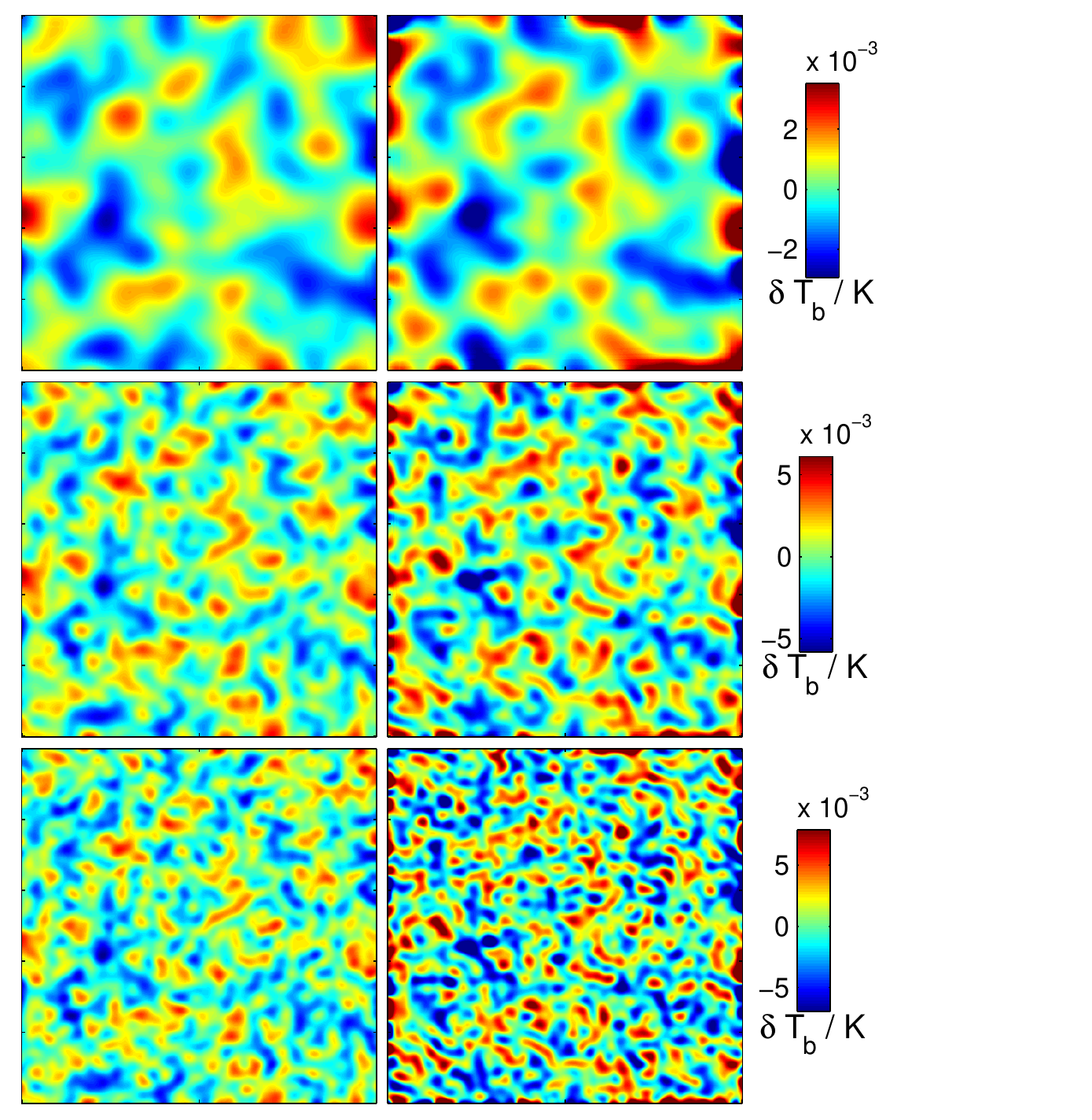}
\caption{In the left column we show the 21-cm signal and in the right column the residuals of \textsc{gmca} at 165 MHz. In the top row only distance scales between 1734 and 108 Mpc are included, in the middle only distance scales between 1734 and 54 Mpc are included and on the bottom the images with all scales present have been smoothed with a 57 Mpc ($\approx$ 20 arcminutes) Gaussian kernel. Clear correlations can be seen between the columns (coefficients of 0.689, 0.687 and 0.588 for the top, middle and bottom rows respectively). Considering only the pixels in a central patch covering 50$\%$ of the total map area we find correlation coefficients of 0.905, 0.788 and 0.605 for the top, middle and bottom rows respectively.}
\label{2map_comp}
\end{figure}

In Fig. \ref{2map_comp} we recover impressive images of the reionization signal at 165 MHz when the smallest scale information is discarded. Comparing the residual and 21-cm maps on each row we find correlation coefficients of 0.689, 0.687 and 0.588 for the top, middle and bottom rows respectively. We therefore conclude that the wavelet decomposition more optimally removes the noise from the residuals than the smoothing technique (bottom row) employed by \citet{zaroubi12}. In Fig. \ref{csfgno_decomp} we can see the wavelet decomposition has the effect of clustering the noise around the edges of the image (this effect is particularly pronounced in rows 3 and 4). To avoid this edge effect we also correlate the maps in Fig. \ref{2map_comp} again but this time considering only the pixels in a central patch covering 50$\%$ of the total map area. We find correlation coefficients of 0.905, 0.788 and 0.605 for the top, middle and bottom rows respectively. With real data, we can assume that this edge effect would not be a problem as the fitting and decomposition can always be carried out over a slightly larger cube with only the central region being used for data analysis.

\section{Conclusions}
\label{conclusions}
In this paper we have assessed the sparsity-based blind source separation technique \textsc{gmca} as a possible way of removing the foregrounds on a 21-cm EoR signal. We recover the 1D, 2D and 3D power spectra to high accuracy across the frequency range. Since the mixing matrix calculated by \textsc{gmca} was shown not to be a strong function of the noise realization, we were able to compensate for leakage of noise power into the reconstructed foregrounds using an independent noise realization, leading to more complete 2D and 3D power spectra.

We also considered if images of reionization could be recovered from the LOFAR-EoR data once foreground removal with \textsc{gmca} has been carried out. Using wavelet decomposition, we considered the phase correlation between the \textsc{gmca} residuals and the simulated 21-cm at different scales. We find strong correlations at the cruder wavelet scales and add several scales together to balance the amount of information in an image with the accuracy of the phase recovery. We find that when distance scales of below 54 Mpc are discounted, the \textsc{gmca} residuals images are highly correlated with the 21-cm images, with correlation coefficients of just less than 0.7. In comparison, smoothing the images did not produce as strong a correlation.

\textsc{gmca} is a highly adaptable method and there remains the possibility that with careful tuning, the 21-cm signal could be picked out as a separate source as opposed to being present as a residual of the process. We intend to explore this further and consider using different mixing matrices for each scale.

\section{Acknowledgments}
FBA acknowledges the support of the Royal Society via a University Research Fellowship. 

GH is a member of the LUNAR consortium, which is funded by the NASA Lunar Science Institute (via Cooperative Agreement NNA09DB30A) to investigate concepts for astrophysical observatories on the Moon.

LVEK acknowledges the financial support from the European Research Council under ERC-Starting Grant FIRSTLIGHT - 258942.

The authors would like to acknowledge Mario Santos for useful discussion.

\bibliography{GMCA_3.0}
\bibliographystyle{mn2e}  

\end{document}